\documentstyle[12pt,epsf,ws]{article}
\newdimen\rotdimen
\def\vspec#1{\special{ps:#1}}
\def\rotstart#1{\vspec{gsave currentpoint currentpoint translate
   #1 neg exch neg exch translate}}
\def\rotfinish{\vspec{currentpoint grestore moveto}}
\def\rotr#1{\rotdimen=\ht#1\advance\rotdimen by\dp#1%
   \hbox to\rotdimen{\hskip\ht#1\vbox to\wd#1{\rotstart{90 rotate}%
   \box#1\vss}\hss}\rotfinish}
\def\rotl#1{\rotdimen=\ht#1\advance\rotdimen by\dp#1%
   \hbox to\rotdimen{\vbox to\wd#1{\vskip\wd#1\rotstart{270 rotate}%
   \box#1\vss}\hss}\rotfinish}%
\def\rotu#1{\rotdimen=\ht#1\advance\rotdimen by\dp#1%
   \hbox to\wd#1{\hskip\wd#1\vbox to\rotdimen{\vskip\rotdimen
   \rotstart{-1 dup scale}\box#1\vss}\hss}\rotfinish}%
\def\rotf#1{\hbox to\wd#1{\hskip\wd#1\rotstart{-1 1 scale}%
   \box#1\hss}\rotfinish}%

\begin{document}

\baselineskip 13 pt

\title{ On the flavour structure of the constituent quark }

\author {Antoni Szczurek$^1$,
Alfons J. Buchmann$^2$, and Amand Faessler$^2$   \\
$^1$ Institute of Nuclear Physics, ul. Radzikowskiego 152,
PL-31-342 Cracow, Poland \\
$^2$ Institut f\"ur Theoretische Physik, Universit\"at T\"ubingen, \\
Auf der Morgenstelle 14, D-72076 T\"ubingen, Germany  }

\maketitle
\begin{abstract}
\baselineskip 13 pt
We discuss the dressing of constituent quarks with
a pseudoscalar meson cloud within the effective chiral quark model.
$SU(3)_f$ symmetry breaking effects are included explicitly.
Our results are compared with those of the traditional
meson cloud approach in which pions are coupled to the nucleon.
The pionic dressing of the constituent quarks explains
the experimentally observed violation of the Gottfried Sum Rule
and leads to an enhanced nonperturbative sea of $q\bar q$ pairs
in the constituent quark and consequently in the nucleon. We find
2.5 times more pions and 10-15 times more kaons in the nucleon than
in the traditional picture.
The $\bar d - \bar u$ asymmetry obtained here is concentrated at
somewhat smaller $x$ and the $\bar u / \bar d$ ratio is somewhat
different than in the traditional meson cloud model of the nucleon.
\end{abstract}

\vskip 1 true cm

The deviation of the Gottfried Sum Rule (GSR) from its
classical value
\cite{G67}
\begin{equation}
S_{G} = \int_{0}^{1} [ F_{2}^{p}(x) - F_{2}^{n}(x) ] \frac{dx}{x}
            = {1\over 3}
\label{GSR}
\end{equation}
observed by the New Muon Collaboration (NMC) at CERN \cite{A91,A'94}
($S_G$ = 0.235 $\pm$ 0.026) has created a lot of interest in the
possible sources of its violation.
It is commonly believed that this
violation is a consequence of an internal asymmetry of
the $\overline{d}(x)$ and $\overline{u}(x)$ quark distributions
in the nucleon.
Since at large $Q^2$ the perturbative QCD evolution is
flavour independent and, to leading order in $\log Q^2$, generates
an equal number of $\overline{u} u$ and $\overline{d} d$ sea quarks
\cite{AP77,RS79} nonperturbative effects must play an important role
here. An asymmetry has been predicted by meson cloud models
in which the physical nucleon contains an admixture of
the $\pi N$ and $\pi \Delta$, etc. components in the Fock expansion
\cite{MCM}.
It has only recently been shown that such a model is consistent with both
the violation of the Gottfried Sum Rule
and with the result of the NA51 CERN experiment on Drell-Yan processes
\cite{HSS95}.

Parallel to the traditional approach
Eichten-Hinchliffe-Quigg \cite{EHQ92} have pointed out
that the effective chiral quark theory formulated by Manohar
and Georgi \cite{MG84} may provide an alternative explanation.
In chiral quark theory, the relevant degrees of freedom
are constituent quarks, gluons, and Goldstone bosons.
The chiral quark model employing both gluon and pion
exchange between constituent quarks together with
corresponding exchange currents, has been fairly
successful in simultaneously explaining the positive parity mass spectrum
and the low-energy electromagnetic properties of the nucleon \cite{BHY}.
It has also been succesfully applied to the two-baryon sector \cite{FAE93}.
Recently, it has been argued \cite{GR95} that
Goldstone boson exchange alone can explain the baryon spectrum without
introducing gluon degrees of freedom. The latter are the main ingredients
of the Isgur-Karl model \cite{IK}.
The important question of the relevant degrees of freedom and
the related question whether the pions couple effectively to the nucleon
or to the constituent quarks is presently actively discussed \cite{EHQ92,GR95}.
At present, it seems premature to decide which picture
of the nucleon is closer to reality and which are the correct degrees
of freedom. Instead it is necessary to study the consequences of
these different scenarios in a broad range of physical processes.

In this paper, we study the flavour structure of
the constituent quark and the nucleon.
While the problem of the flavour structure of the nucleon,
and the $\bar d - \bar u$ asymmetry
has been recently discussed in some detail within the conventional
mesonic cloud picture \cite{HSS95},
no detailed analysis exists in the chiral quark model ($\chi$QM).
In Ref.\cite{EHQ92} the $\chi$QM was used
as a motivation to introduce $SU(2)$ asymmetric parametrizations for
the $x$ dependence of the $\bar d $ and $\bar u$ distributions.
In this work we calculate the $\bar d - \bar u$ asymmetry directly from the
$\chi$QM.
In particular, we discuss the effect of $SU(3)_f$
symmetry breaking which was not considered in Ref.\cite{EHQ92}.
This may be especially important in understanding the strangeness
content of the nucleon and the closely related nucleon spin problem \cite{A89}.
We also study the implications of the GSR violation on the
$\Delta -N$ mass splitting.

The interaction Lagrangian of the effective chiral
quark theory \cite{MG84} is in leading order of
an expansion in $\Pi/f$
%
\begin{equation}
\label{chiral}
{\cal L}_{int} = - {g_A \over f}
\Psi \partial_{\mu} \Pi \gamma^{\mu}\gamma_5 \Psi \,
\end{equation}
where $\Pi$ is the Goldstone boson field, $f \approx 93$ MeV
the pion decay constant,
and $\Psi$ the constituent quark field.
The effective chiral Lagrangian of Eq.(\ref{chiral}) describes the
coupling
of Goldstone bosons to massive ($m_Q \approx m_N/3$) constituent quarks.
Both the mass of the constituent quark and its coupling to
Goldstone bosons are consequences of
the spontaneously broken chiral symmetry of QCD.
The light-front Fock decomposition of the constituent quark wave
functions (see also Fig.1) reads
\begin{eqnarray}
|U \rangle &=& Z^{1/2} |u \rangle +
\sqrt{1 \over 3} \alpha_{\pi/U} |u \pi^0 \rangle +
\sqrt{2 \over 3} \alpha_{\pi/U} |d \pi^+ \rangle +
\alpha_{K/U} | s K^+ \rangle + ... \; ,
\label{decomposition}
\nonumber \\
|D \rangle &=& Z^{1/2} |d \rangle +
\sqrt{1 \over 3} \alpha_{\pi/D} |d \pi^0 \rangle +
\sqrt{2 \over 3} \alpha_{\pi/D} |u \pi^- \rangle +
\alpha_{K/D} | s K^0 \rangle + ... \; ,
\end{eqnarray}
where capital (small) letters denote constituent quarks dressed (undressed)
by Goldstone bosons and $Z$ is a wave function renormalization.

For simplicity, we list all formulae for pions although kaons are included
in the actual calculation.
In analogy to the nucleonic Sullivan process \cite{HSS95,S72}
in deep-inelastic scattering (DIS), we
consider the pion-quark splitting function
$f_{q \rightarrow \pi q'}(x_{\pi},k_{\perp}^2)$
(flux factor summed over quark spin
polarizations). The splitting function
determines the probability for finding a Goldstone
boson of mass $m_{\pi}$ carrying the light-cone momentum fraction $x_{\pi}$
of the parent constituent quark $Q$
\begin{equation}
f_{q \rightarrow \pi q'}(x_{\pi},k_{\perp}^2) =
{g_{QQ'\pi}^2 \over 16 \pi^2}
{1 \over x_{\pi} (1-x_{\pi})}
|G_{QQ'\pi}(x_{\pi}, k_{\perp}^2)|^2
\frac{((1-x_{\pi})m_{Q} - m_{Q'})^2 + k_{\perp}^2}
{(1-x_{\pi}) (m_{Q}^2 - M^2_{\pi Q'})^2},
\end{equation}
where $k_{\perp}$ is the perpendicular momentum of the recoiling quark $q'$.

The constituent quark-pion coupling constant can be obtained
from the quark version of the Goldberger-Treiman relation
\begin{equation}
g_{QQ'\pi^{0}}^2 = {g_A^2 \over f^2} {(m_Q + m_{Q'})^{2} \over 4} \, ,
\end{equation}
with $g_A$ being the axial-vector constant of the constituent quark.
In practical calculations we investigate two cases:
$g_A = 1$ as suggested by an $1/N_c$
expansion \cite{W90} (model A); and $g_A = 0.75 $ as suggested by the
nonrelativistic quark model \cite{MG84} (model B).
We also take
$m_l = m_{Q} = m_{Q'} = m_N / 3 =$
313 MeV for the light {\it up} and {\it down} quarks and
$m_s = m_{Q'} = m_{\Sigma} - m_N + m_{l} =$
567 MeV for the {\it strange} quarks. As in ref. \cite{EHQ92} we do not
explicitely calculate the contribution of the meson cloud
to the mass and coupling constant of the constituent quark
but consider all masses and coupling constants
as renormalized quantities.

The $G_{QQ'\pi}(x_{\pi}, k_{\perp}^2)$ is a vertex function, which
accounts for the extended structure of both the pion
(and other Goldstone bosons) and the constituent quark
\begin{equation}
G_{QQ'\pi}(x_{\pi}, k_{\perp}^2) =
\exp \left(\frac{m_Q^2 - M^2_{\pi Q'}
(x_{\pi}, k_{\perp}^2)}{2 \Lambda^2} \right),
\label{vff}
\end{equation}
with $M^2_{\pi Q'}(x_{\pi}, k_{\perp}^2) =
{ (m_{\pi}^2 + k_{\perp}^2)/ x_{\pi} } +
{ (m_{Q'}^2 + k_{\perp}^2 )/( 1-x_{\pi}) } $,
being the invariant mass squared of the $\pi + Q'$
system. A form factor of this type fulfills the number and momentum sum rules
by construction \cite{HSS95}.

Isospin symmetry leads to the following simple relations for the
integrated (over $k_{\perp}^2$) pion and kaon
splitting functions
\begin{eqnarray}
f_{u\rightarrow \pi^+ d}(x_{\pi}) =
f_{d\rightarrow \pi^- u}(x_{\pi}) =
2 \, f_{u\rightarrow \pi^0 u}(x_{\pi}) =
2 \, f_{d\rightarrow \pi^0 d}(x_{\pi}) \; ,
\nonumber \\
f_{u\rightarrow K^+ s}(x_K) = f_{d\rightarrow K^0 s}(x_K).
\end{eqnarray}
The integral of the splitting function
\begin{equation}
P_{M/Q} = |\alpha_{M/Q}|^2 =
\sum_{q'} \int_0^1 f_{q \rightarrow M q'}(x_{M}) \, dx_{M} \; ,
\end{equation}
is the probability of finding a Goldstone boson $M$ in
the constituent quark $Q$ and $\alpha_{M/Q}$ is the corresponding amplitude
appearing in Eq.(\ref{decomposition}).

The regularization parameter $\Lambda$ in (\ref{vff}) is not known
a priori.
Assuming that the GSR violation
\begin{equation}
S_G = {1 \over 3} + {2 \over 3}
\int_0^1 \left( \bar u(x) - \bar d(x) \right ) dx =
{1 \over 3} - {4 \over 9} P_{\pi/Q}
\end{equation}
is entirely due to the dressing of the constituent quark by pions
one obtains $\Lambda$ by fitting $P_{\pi/Q}$ to the NMC value for
$S_G$ \cite{A'94}.
In Fig.2a we present the total splitting function of
the constituent quark
$f_{M/Q}(x_M) \equiv \sum_{q'} f_{q\rightarrow M q'}(x_{M})$
into the pion $M=\pi$ and kaon $M=K$ for model A (solid line)
and model B (dashed line).
The average momentum fraction carried by the meson in the
$|M Q'\rangle$ Fock state is
$\langle x_{\pi} \rangle$ = 0.597 (A), 0.594 (B) and
$\langle x_{K} \rangle$ = 0.606 (A), 0.589 (B) for
the pion and kaon, respectively.

By construction the number of
pions, $P_{\pi/Q}$ = 0.22 (0.22), remains the same, but the number of kaons
$P_{K/Q}$ = 0.051 (0.084) is different for models A(B).
Thus, the number of pions and kaons in the nucleon is respectively
$P_{\pi/N}$ = 0.66
and $P_{K/N}$ = 0.15-0.25. These numbers are considerably larger
than those found in traditional nucleonic meson cloud models
\cite{MCM}.
%
%
We get considerable damping of the kaon
splitting function with respect to the pion splitting function.
The strong suppression of the kaonic loops with respect to pionic loops
is caused by
the large mass difference between kaons and pions
and between strange and non-strange constituent quarks.
Due to the inclusion of these $SU(3)_f$ symmetry breaking
effects we find a considerably smaller number of strange quarks in the
nucleon than EHQ \cite{EHQ92}
( 0.15-0.25 here vs. 0.63 in EHQ ).
However, our result for the number of strange quarks
is still a factor of 10-15(!) larger in comparison to
the traditional meson cloud model \cite{HSS95}.

These results have direct
consequences for the spin problem. Assuming a naive $SU(6)$
spin-flavour constituent quark wave function of the nucleon we get
an upper limit for the strange quark contribution to the nucleon
polarization
\begin{equation}
|\Delta s_N| = |\Delta s_Q| \, < \,
P_{s/Q} = P_{K/Q} = 0.051 (A), 0.084 (B)
\end{equation}
and a lower limit for the spin polarization carried by quarks
\begin{equation}
1 \, > \Sigma \, > { 1 - 2(P_{\pi/Q} + P_{K/Q})}
 = 0.46 (A), 0.39 (B) \; .
\end{equation}

As a direct consequence of the pion cloud dressing, the
constituent $U$ and $D$ quarks
in the proton ($UUD$) and neutron ($DDU$)
contain not only $up$ and $down$ quarks, respectively, but also
some admixture of (anti)quarks of different flavours.
Formally, the DIS-quark distributions in the constituent $U$ or $D$
quarks at the initial scale of the QCD evolution can be written as
%
%
%
\begin{equation}
u_U(x)  =  u_U^{(0)}(x) + u_U^{(i)}(x) + u_U^{(\pi)}(x)
\; , \qquad d_D(x)  =  d_D^{(0)}(x) + d_D^{(i)}(x) + d_D^{(\pi)}(x) \; ,
\end{equation}
\begin{equation}
d_U(x)  =                d_U^{(i)}(x) + d_U^{(\pi)}(x)
\; , \qquad u_D(x)  =                 u_D^{(i)}(x) + u_D^{(\pi)}(x) \; ,
\end{equation}
where the contributions denoted with (0) correspond to the bare
(undressed of pions) constituent quarks, those denoted with (i) to
the intermediate quarks associated with pions and finally those
denoted with $(\pi)$ originate from the pion.
The distribution of the bare (undressed of pions) quarks in the
constituent quarks is
\begin{equation}
u_U^{(0)}(x) = d_D^{(0)}(x) = (1- \sum_M P_{M/Q})
\; \delta(x-1) .
\end{equation}
The contribution of the $Q'$ (intermediate) quarks
is fully determined by the pion splitting function
\begin{equation}
u_U^{(i)}(x) = d_D^{(i)}(x) = {1 \over 3} f_{\pi/Q}(1-x) \; ,
\qquad
u_D^{(i)}(x) = d_U^{(i)}(x) = {2 \over 3} f_{\pi/Q}(1-x) \; .
\end{equation}

We assume hereafter that at the confinement scale,
antiquarks originate exclusively from the virtual Goldstone bosons.
In analogy to the classical Sullivan process \cite{S72},
the antiquark distributions can be calculated as
\begin{eqnarray}
\bar u_U(x) = \bar u_U^{\pi}(x) = \bar d_D(x) = \bar d_D^{(\pi)}(x)
={1 \over 6} I_{\pi}(x)  \; , \nonumber \\
\bar u_D(x) = \bar u_D^{\pi}(x) = \bar d_U(x) = \bar d_U^{(\pi)}(x)
= {5 \over 6} I_{\pi}(x) \; ,
\end{eqnarray}
where
$I_{\pi}(x) = \int_x^1 dy \, y^{-1} f_{\pi/Q}(y) q_{\pi}(x/y) \, .$

As an example, we show in Fig.2b the $x \bar u(x)$ (solid),
$x \bar d(x)$ (dashed) and $x (s(x) + \bar s(x)) / 2$ (dotted)
DIS-quark distributions in the constituent $U$ quark at the initial
QCD scale. By an appropriate isospin rotation
corresponding distributions are obtained inside the constituent $D$ quark.
We find a large asymmetry between $\bar d$ and $\bar u$ quark distributions
and a rather large (anti)strange quark component.
This will have important consequences for the nucleon sea.
In this calculation we have taken the quark distributions in the pion
as parametrized for different values of $Q^2$ in Ref.\cite{GRV92_pi},
where they have been adjusted to describe the pion-nucleus Drell-Yan data.

The quark distributions in the nucleon $q_{f,N}(x)$
can be obtained from those of the constituent quarks as
\begin{equation}
q_{f,N}(x) = \int_x^1 \biggl [
U_N(y) q_{f,U}(x/y) + D_N(y) q_{f,D}(x/y)
\biggr ] \, {dy \over y} \; .
\end{equation}
The consistency of our approach
requires that the distributions of the constituent quarks
$U_N(y)$ and $D_N(y)$ inside the nucleon are $Q^2$ independent
in contrast to
$u_U(x,Q^2)$, $u_D(x,Q^2)$, $d_U(x,Q^2)$, $d_D(x,Q^2)$, etc. which
are subjected to the QCD evolution.
In practical calculations we parametrize the distributions of
constituent quarks in the nucleon as
$Q_N(y) = C_{\alpha \beta} \, y^{\alpha}(1-y)^{\beta}$.
The parameters $\alpha$ and $\beta$ can be obtained from
the requirements
$\int_0^1 Q_N(y) \, dy = 1$ (number sum rule) and
$ 3 \int_0^1 y \, Q_N(y) \, dy = 3/4$ (momentum sum rule)
and by imposing the counting rules at $y \rightarrow 1$.
This yields $\alpha = 1/3$ and $\beta = 3$.
The number and momentum sum rules put stringent constraints on
the quark distributions in any model.
Following Ref.\cite{GRV95} we assume a {\it valence-like} gluon
distribution which for simplicity is taken to be identical to the valence
quark distribution in the nucleon $g(x,Q_0^2) = Q_N(x)$.
Fairly similar gluon distributions can be obtained by dressing quarks
with gluons in the nonperturbative regime with massive ($m_g^{eff}$)
effective gluons and frozen running $\alpha_s$. Rather heavy effective
gluons $m_g^{eff} >$ 0.4 GeV and small $\alpha_s < 0.5 $ are required
in order to limit the momentum carried by quarks to approximately 1/4
as required by the phenomenology \cite{GRV95}.

In Fig.3a we compare the $\chi$QM prediction for
the antiquark distributions $x \bar u(x)$ and $x \bar d(x)$ in the proton
to the phenomenological GRV antiquark distributions at low
momentum transfers \cite{GRV95}.
The $\chi$QM antiquark distributions peak approximately at the same
Bjorken-x but are considerably smaller.

In comparison to the traditional formulation of the meson cloud model
\cite{HSS95} the strange sea quark distributions predicted by the
$\chi$QM (shown in Fig.3b) are enhanced.
Similar to the traditional nucleonic meson cloud approach \cite{HSS95}
we get $s(x) \ne \bar s(x)$. In contrast to the nucleonic meson cloud
picture, the quark meson cloud approach leads to some difference between
$s(x)$ and $\bar s(x)$ distributions which could be detected in the
(anti)neutrino DIS experiments.
The momentum carried by the sea quarks
$\sum_f \int_0^1 x \left (q_f^{sea}(x) + \bar q_f^{sea}(x) \right) dx =
2 \sum_f \int_0^1 x \bar q_f(x) dx$ = 0.08-0.09 This
large number remains, however, nearly unchanged by the QCD evolution
and is somewhat smaller than the result
of the CCFR collaboration \cite{F90} at $Q^2$ = 16.85 GeV$^2$.

In Fig.4a we compare
the antiquark distributions at $Q^2$ = 4 GeV$^2$, obtained
from the ones of Fig.3 by QCD evolution \cite{AP77}, to the recent
Martin-Roberts-Stirling (MRS A) parametrization of the world data
on DIS and Drell-Yan processes \cite{MSR94}.
The leading order (LO) anti-quark distributions obtained in the $\chi$QM
are significantly smaller than the quark distributions obtained from
the next-to-leading order (NLO) analysis of Martin, Roberts and Stirling
\cite{MSR94}.
Considerable part of this effect is due to the known difference
between LO and NLO antiquark distributions (for an illustration see
\cite{BAB95}).
A big fraction of the missing strength is presumably due to the neglect
of the quark meson exchange currents \cite{BHY}.
This deserves further study in the future.
The effect of the meson exchange currents cancels in the difference
$x(\bar d - \bar u)$ which is shown in panel (b).
In comparison to the MRS(A) parametrization and the traditional
meson cloud approach \cite{HSS95}, the $\chi$QM result for this difference
is concentrated at smaller Bjorken-$x$.
In panel (c) we present (anti)strange quark distributions obtained from
our model at Q$^2$ = 4 GeV$^2$.

While at present the extraction of the $x$-dependence of various sea
quark components is a matter of some controversy, the total sea quark
distribution $x \bar q (x) = x (\bar u (x) + \bar d (x) + \bar s (x))$
can be obtained from the (anti)neutrino DIS \cite{M92}.
In Fig.4d we confront the $x$-dependence obtained from the chiral quark
model with the experimental data of the CCFR collaboration
at Q$^2$ = 3 and 5 GeV$^2$.
The antiquark distribution $x \bar q (x)$ obtained in the $\chi$QM
underestimates the experimental data by about 20-30\%, leaving room
for some other unknown contributions. We expect the meson exchange
current contribution to be the dominant missing contribution.

It is instructive to study different
ratios of the quark distributions rather than the quark distributions
themselves. In Fig.4e we present the ratio
$R(x) \equiv \bar u(x) / \bar d(x)$
and in Fig.4f the ratio
$R_s(x) \equiv \frac{s(x) + \bar s(x)}{\bar u(x) + \bar d(x)}$.
The latter is usually {\it assumed} to be a constant in all
available parametrizations of the data (including MRS(A)).
The simple model discussed here predicts an interesting Bjorken-$x$
dependence of $R_s(x)$ which could be the subject of a
dedicated experimental study.
In Figs. 4(e-f) we show also the corresponding ratios
at the inital confinement scale $Q_0^2$.
The ratio $R(x,Q_0^2) = \bar u(x,Q_0^2) / \bar d(x,Q_0^2)$
(dashed line) is independent of Bjorken-$x$ and equals
to $7 \over 11$. Since the QCD evolution (solid line) even enhances
this ratio, our result for $R(x,Q^2)\approx 0.74$
is too large compared to the recent NA51 CERN experiment \cite{B94}
$R=\bar u / \bar d = 0.51 \pm 0.04 (stat) \pm 0.05 (syst) $
at $x = 0.18$. Note that our prediction for the $x$-dependence of this ratio
is quite different from the MRS(A) parametrization. A measurement of
$R(x)$ \cite{G92} would shed further light on the problem,
which picture (traditional meson cloud vs. chiral quark model)
is more appropriate.
%

Finally, we study the consequences of the $\chi$QM
for the $N -\Delta$ mass splitting $\delta^{N\Delta}\equiv m_{\Delta}-m_N$.
Both, the spin-dependent gluon and pion exchange potentials between
constituent quarks contribute to $\delta^{N\Delta}=
\delta_g^{N\Delta} + \delta_{\pi}^{N\Delta}$.
The size of the pion contribution $\delta_{\pi}^{N\Delta}$ is
mainly determined by the (unknown) structure of the $QQ'\pi$
vertex and is therefore model-dependent \cite{BHY,GR95}.
Fixing the cut-off parameter $\Lambda$ of the $QQ'\pi$ vertex
by the experimental value for the Gottfried sum rule, $S_G$,
also fixes $\delta_{\pi}^{N\Delta}$.
We calculate $\delta_{\pi}^{N\Delta}$ for both models A and B using
$G_{QQ'\pi}(t) =\left( \frac{\Lambda_{}^2}{\Lambda_{}^2-t} \right)^{1/2}$
\cite{BHY}
and determine $\Lambda$ from the experimental value of $S_G$.
We obtain for model A:
$\Lambda$ = 1.26   GeV  which corresponds to
$\delta_{\pi}^{N\Delta}$ = 222   MeV.
Likewise we obtain for model B:
$\Lambda $ = 3.31  GeV
and $\delta_{\pi}^{N\Delta}$ = 140  MeV.
Evidently, $\delta_{\pi}^{N\Delta}$
depends strongly on the
pion-quark coupling constant $g_{\pi QQ'}$
for which quite different values have been used in the
recent literature
\cite{EHQ92,BHY,BPKP95}. However, even in the extreme case of a very
strong $g_{\pi QQ'}$ (model A) we obtain only about $3/4$
of the experimental $\delta^{N\Delta}$.
We have checked that this conclusion does not depend
on the functional form of the $QQ'\pi$ vertex.

%

Summarizing, we have studied the flavour structure of the nucleon
in the effective chiral quark model \cite{MG84} in which
the Goldstone bosons couple directly to the constituent quarks.
With a {Goldstone boson -- constituent quark}
light-cone wave function adjusted to reproduce the experimental Gottfried
Sum Rule \cite{A'94}, we have calculated the resulting antiquark
distributions inside the constituent quark and inside the nucleon.
We find 2--3 times more pions and 10--15 times more kaons in the nucleon
than in the traditional meson cloud model in which the Goldstone bosons
couple effectively to the nucleon\cite{HSS95}. In general, the
corresponding sea is concentrated at rather small Bjorken-x.
The predicted $\bar u(x) / \bar d(x)$ ratio is larger than the one
obtained by the NA51 experiment at CERN \cite{B94}.
It may be expected that the gluon-exchange interaction between
constituent quarks, which leads to the different x-dependence
of up and down valence quarks, may to some extent modify the
$\bar u(x) / \bar d(x)$ ratio obtained here.
Additional measurements of the $x$ dependence of
this ratio are required to distinguish between different models.
In comparison to the nucleonic meson cloud model, the
$x (\bar d - \bar u)$ difference is concentrated at smaller Bjorken-x,
rather inconsistent with the recent MRS phenomenological analysis
\cite{MSR94} (see also a discussion in Ref.\cite{K95}).
The discrepancy with the Drell-Yan data and the phenomenological MRS
analysis may, in our opinion, be due to the many-body effects
neglected in independent dressing of (interacting) constituent quarks.
These effects are rather difficult to include on the microscopic level.
In the traditional (nucleonic) formulation of the meson cloud
they are treated in the strong binding limit
(see a discussion in Ref.\cite{JM91}).
The $\chi$QM leads to enhanced strange sea distributions and a
measurable difference between $s(x)$ and $\bar s (x)$ distributions.
Finally, the experimental Gottfried Sum Rule violation provides
stringent limits on the pionic contribution to the nucleon-delta
mass splitting.

\baselineskip 12pt


\vfill
\eject

\pagestyle{empty}

\centerline{ {\bf Figures} }

\vspace{4.5cm}

\begin{figure}[htb]
$$\vspace{2.5 cm} \hspace{0.2cm} \mbox{
\epsfxsize 5.5 true cm
\epsfysize 16.0 true cm
\setbox0= \vbox{
\hbox {
\epsfbox{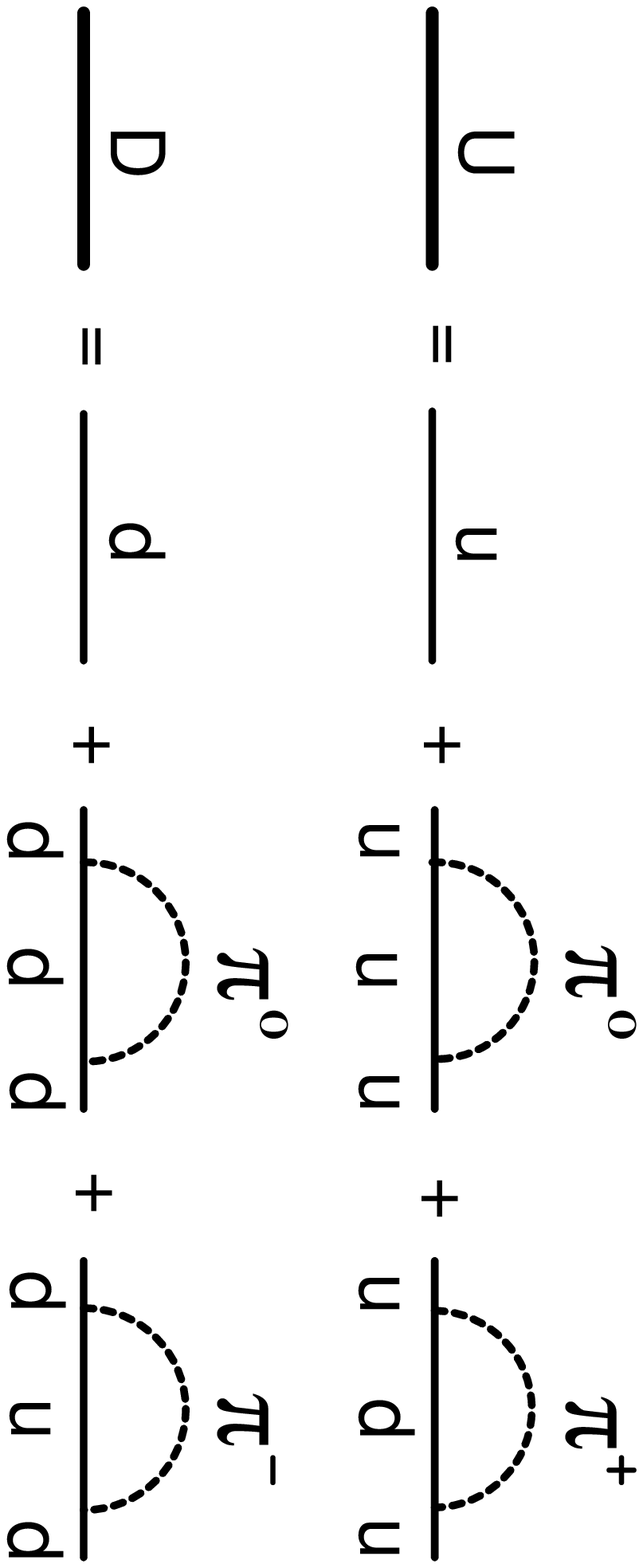}
} 
} 
\rotl0
} $$
\vspace{0.5cm}
\caption[Fig.1]{
The dressing of the constituent quarks with pions.}
\end{figure}

\vfill
\eject

\begin{figure}[htb]
$$\hspace{0.2cm} \mbox{
\epsfxsize 17.0 true cm
\epsfysize 23.0 true cm
\setbox0= \vbox{
\hbox {
\epsfbox{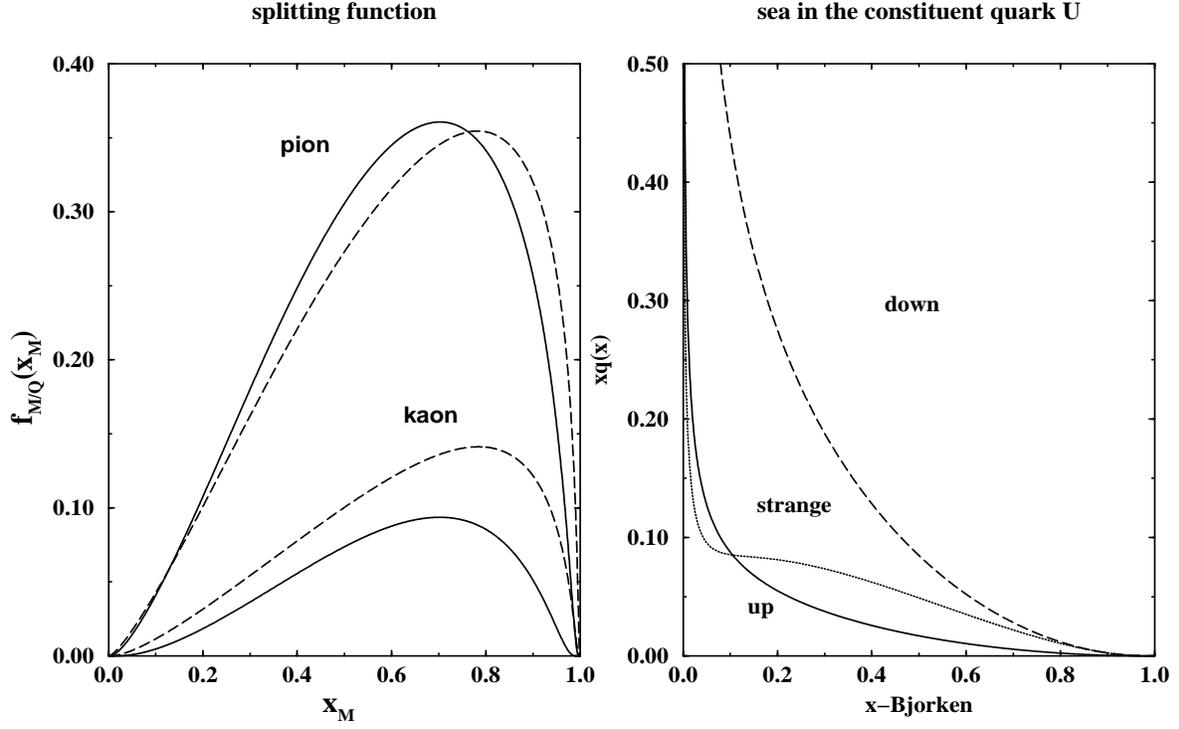}
} 
} 
\box0
} $$
\vspace{-4.5cm}

\caption[Fig.2]{
(a) Total splitting function (flux factor) of the constituent quark into
the pion and kaon as a function
of the light-cone momentum fraction $x_M$ carried by the Goldstone
boson in the constituent quark; model A (solid lines) and model B
(dashed lines). The corresponding vertex function parameters
of the light-cone wave function $|MQ' \rangle$,
$\Lambda$=2.287 GeV (A)  and 5.5 GeV (B)
have been obtained
by fitting to the experimental value of $S_G$ \protect\cite{A'94}.
(b) The $x \bar u(x)$ (solid), $x \bar d(x)$ (dashed) and
$x (s(x) + \bar s(x)) / 2$ (dotted) DIS-quark distributions
in the constituent  $U$  quark at the initial low-momentum scale.}
\end{figure}

\vfill
\eject

\begin{figure}[htb]
$$\hspace{0.2cm} \mbox{
\epsfxsize 17.0 true cm
\epsfysize 23.0 true cm
\setbox0= \vbox{
\hbox {
\epsfbox{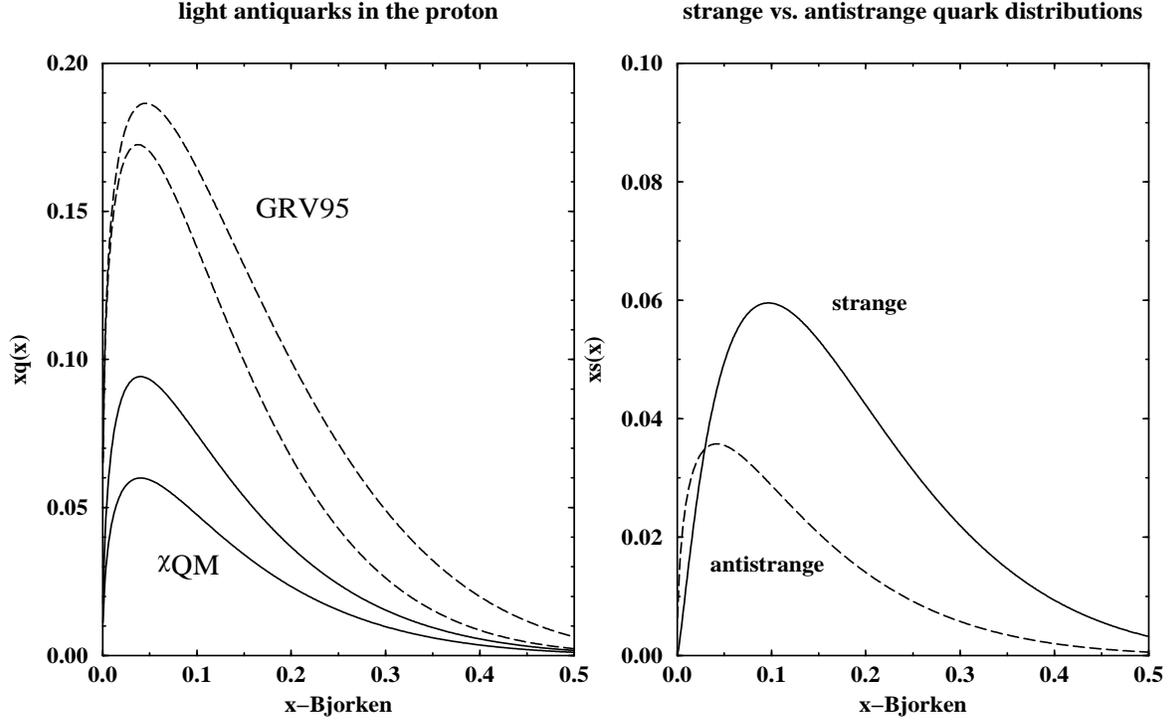}
} 
} 
\box0
} $$
\vspace{-4.5cm}
\caption[Fig.3]{
Antiquark momentum distributions in the nucleon
for the $\chi$QM model A.
(a) $x \bar u(x)$ and $x \bar d(x)$
antiquark distributions in the proton (solid lines).
For comparison we show the phenomenological antiquark distribution
used in Ref.\protect\cite{GRV95} (dashed lines, GRV95).
Note that the $\bar d$ distributions are always above the $\bar u$
distributions.
(b) $x s(x)$ (solid) and $x \bar s(x)$ (dashed).}
\end{figure}

\vfill
\eject

\vspace{-0.75cm}
\begin{figure}[htb]
\label{Fig.4}
$$\hspace{0.2cm} \mbox{
\epsfxsize 17.0 true cm
\epsfysize 23.0 true cm
\setbox0= \vbox{
\hbox {
\epsfbox{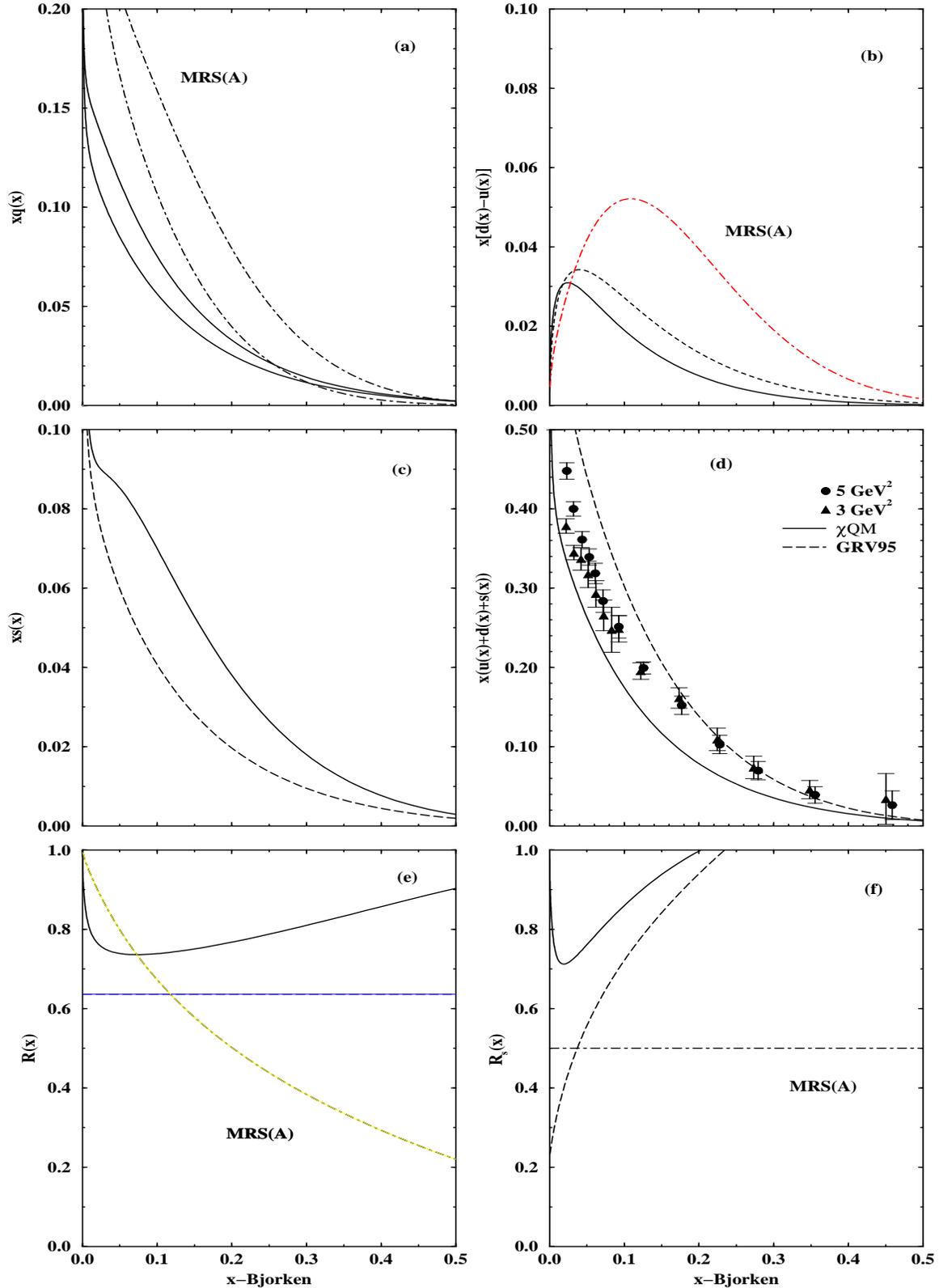}
} 
} 
\box0
} $$
\vspace{-1.3cm}
\caption[Fig.4]{
Antiquark momentum distributions in the nucleon at $Q^2$ = 4 GeV$^2$
(solid lines) as calculated from the ones of Fig.3 by QCD evolution.
The results at the initial scale $Q_0^2=0.25$ GeV$^2$,
are shown by the dashed lines.
Here, $\Lambda_{QCD} = 200 MeV$ and
the number of active quark flavours is $n_f = 3$.
For comparison we show the recent MRS(A) parametrization
\protect\cite{MSR94} (dashed-dotted line)
(a) $x \bar u(x)$ and $x \bar d(x)$,
(b) $x (\bar d(x) - \bar u(x) )$,
(c) $x s(x)$ (solid) and $x \bar s(x)$ (dashed),
(d) $x (\bar u(x) + \bar d(x) + \bar s(x)) $ compared
with the experimental data of the CCFR collaboration \protect\cite{M92},
(e) $R(x) = \frac{\bar u(x)}{\bar d(x)}$,
(f) $R_s(x) = \frac{s(x) + \bar s(x)}{\bar u(x) + \bar d(x)}$.}
\end{figure}

\end{document}